\documentclass[twocolumn,nameyear,amsthm]{autart}

\usepackage{amsmath,amssymb,mathrsfs}
\usepackage{booktabs}
\usepackage{tikz}
\usepackage{pgfplots}
\pgfplotsset{compat=1.18}
\usetikzlibrary{arrows.meta}
\usepackage{float}
\usepackage{enumitem}
\usepackage[authoryear]{natbib}
\usepackage{flushend}

\AtBeginDocument{\begingroup\setbox0=\hbox{$\relax$}\endgroup}

\newtheorem{assumption}{Assumption}
\newtheorem{theorem}{Theorem}
\newtheorem{lemma}{Lemma}
\newtheorem{proposition}{Proposition}
\newtheorem{corollary}{Corollary}

\date{31 October 2026}

\begin{document}

\begin{frontmatter}

\title{Finite Global $H_{\infty}$ Optimization of the $Q$-Filter Time Constant for Nominal-Response Recovery in SISO Disturbance Observers under Parametric Plant Uncertainty}

\author[umd]{Hyuntae Kim}\ead{htkim26@umd.edu}
\and
\author[polyu]{Wen-Hua Chen\corauthref{corinfo}}\ead{wenhua.chen@polyu.edu.hk}
\corauth[corinfo]{Corresponding author.}
\address[umd]{Department of Mechanical Engineering, University of Maryland, 2107 Glenn L. Martin Hall, 4298 Campus Drive, College Park, MD 20742, USA}
\address[polyu]{Department of Aeronautical and Aviation Engineering, The Hong Kong Polytechnic University, Hung Hom, Kowloon, Hong Kong SAR, China}

\begin{abstract}
This paper develops a finite real-algebraic procedure for global H-infinity optimization of the time constant of a fixed-shape Q-filter in a single-input/single-output (SISO) disturbance observer (DOB) under affine parametric plant uncertainty. Robust internal stability defines a possibly disconnected feasible domain, and the objective is the worst-case weighted mismatch between the uncertain DOB reference response and the nominal closed-loop response. A recovery-error factorization preserves affine dependence of the performance numerator and denominator on the same uncertainty vector. Phase lifting and a complex-polynomial Edge theorem reduce the uncertainty box to finitely many polynomial edges, while quadratic dependence of the squared frequency-domain inequality on the edge coordinate permits analytic elimination of that coordinate. The resulting semialgebraic formulation has only the squared frequency as a universally quantified variable. It determines the global infimum, decides whether it is attained, and constructs the complete optimizer set over the full robust-stability domain. A DOB case study with a plant family containing non-minimum-phase members yields a unique optimal time constant of approximately 0.0364075 and a worst-case norm of approximately 2.29975. The optimizer lies strictly inside the first robust-stability component, and every smaller feasible time constant in that component has a larger worst-case norm. Finite algebraic certificates establish robust stability, performance feasibility, and strict global optimality.
\end{abstract}


\end{frontmatter}

\section{Introduction}

Disturbance-observer-based control (DOBC) augments a predesigned nominal controller with a disturbance-estimation and compensation loop. This modular viewpoint is central to the two-stage DOBC framework and its subsequent developments \citep{Chen2004,Chen2016}. In the conventional linear disturbance observer (DOB), a stable low-pass $Q$-filter controls how rapidly the uncertain plant is driven toward the nominal closed-loop behavior; for a fixed normalized filter shape, decreasing the time constant increases the characteristic bandwidth, drives the filter toward unity, and, under the usual high-bandwidth stability conditions, promotes nominal-performance recovery on each fixed finite-frequency interval \citep{ShimJo2009,Shim2016,Chang2020}.

When the uncertainty family contains non-minimum-phase (NMP) plants, time-constant selection must also account for the global robust-stability geometry. Robust internal stability can exclude all sufficiently small values and can produce several disconnected admissible intervals. Moreover, lightly damped modes near stability boundaries can generate narrow peaks that dominate the worst-case frequency response. Consequently, choosing a time constant arbitrarily close to the lower robust-stability boundary need not minimize the worst-case $H_{\infty}$ loss of nominal behavior.

Closed-loop $H_{\infty}$ formulations have previously been used to design fixed-structure disturbance observers. In particular, \citet{WangTomizuka2004} considered richer observer parameters and mixed closed-loop specifications, leading to a nonconvex static-output-feedback problem addressed by a numerical suboptimal design procedure. More generally, fixed-order and structured $H_{\infty}$ design is commonly treated by nonsmooth, nonconvex numerical optimization \citep{ApkarianNoll2006,GumussoyOverton2008}. These approaches accommodate richer parameterizations.

Let $s$ denote the Laplace variable and let $\overline Q$ be a prescribed normalized filter shape. The present work tunes the $Q$-filter time constant $\tau>0$ in the conventional family $Q(s;\tau)=\overline Q(\tau s)$. It solves this scalar problem globally over the complete robust-stability domain, including the attainment decision. Let $\Theta$ denote the affine plant-uncertainty box, let $\mathcal R_{\rm stab}$ denote the full robust-stability domain, and let $W$ be a prescribed stable weight. Let $\theta\in\Theta$ denote the plant-uncertainty vector. Let $T_{yr}(s;\tau,\theta)$ denote the uncertain DOB reference-to-output response and let $T_{\rm n}(s)$ be the response intended by the nominal controller. The design criterion is
\begin{equation}
 \inf_{\tau\in\mathcal R_{\rm stab}}
 \sup_{\theta\in\Theta}
 \left\|W(s)\bigl(T_{yr}(s;\tau,\theta)-T_{\rm n}(s)\bigr)\right\|_{\infty}.
 \label{eq:intro_problem}
\end{equation}
Here, $\|\cdot\|_\infty$ denotes the $H_\infty$ norm. The objective is to determine the global infimum, its attainment status, and every minimizing time constant across all robust-stability components by a finite algebraic computation. We refer to this weighted reference-response measure as nominal-response recovery, a specific nominal-performance-recovery channel.

The main contributions are as follows.
\begin{enumerate}[label=(\roman*),leftmargin=5mm]
\item The DOB recovery error is factorized so that its numerator and denominator retain affine dependence on the same physical uncertainty vector.
\item Phase lifting, a complex-polynomial Edge theorem, and analytic elimination of the edge coordinate reduce the robust $H_{\infty}$ condition from $p+1$ quantified real variables, where $p$ is the uncertainty dimension, to the single squared-frequency variable.
\item The reduced condition yields a finite real-algebraic algorithm that computes the global infimum, decides attainment, and constructs the complete optimizer set over a possibly disconnected robust-stability domain.
\item The algebraic calculation for a DOB case study with NMP plant uncertainty yields a singleton global optimizer over the full robust-stability domain.
\end{enumerate}
The phase-lifting and polynomial-edge theorems are established robust-control tools \citep{Bartlett1988,ShiZhang1995,BialasGora2012,Wang2003}. Combining them with the shared DOB uncertainty structure converts architecture-preserving bandwidth selection into a finite global-optimization computation.

\section{DOB model and global tuning objective}

\subsection{Plant family and conventional DOB equations}

Consider the proper real-rational single-input/single-output (SISO) plant, nominal model, and controller
\begin{equation}
 P(s,\theta)=\frac{N(s,\theta)}{D(s,\theta)},\qquad
 P_{\rm n}(s)=\frac{N_{\rm n}(s)}{D_{\rm n}(s)},
\end{equation}
\begin{equation}
 C(s)=\frac{N_C(s)}{D_C(s)}.
\end{equation}
Each displayed transfer function is represented by a real coprime numerator-denominator pair whose denominator has positive leading coefficient. Let $p\in\mathbb N$ be the uncertainty dimension and assume $\underline\theta_k<\overline\theta_k$ for $k=1,\ldots,p$. The uncertainty vector belongs to the box
\begin{equation}
 \Theta:=\prod_{k=1}^{p}[\underline\theta_k,\overline\theta_k],
\end{equation}
and the plant polynomials have the shared affine representation
\begin{align}
 N(s,\theta)&=N_0(s)+\sum_{k=1}^{p}\theta_kN_k(s),\\
 D(s,\theta)&=D_0(s)+\sum_{k=1}^{p}\theta_kD_k(s).
\end{align}

Let $\zeta$ denote the normalized complex-frequency variable, and let
\begin{equation}
 \overline Q(\zeta)=\frac{\overline N_Q(\zeta)}{\overline D_Q(\zeta)}
\end{equation}
be a fixed stable proper real-rational low-pass filter represented by a real coprime polynomial pair normalized by
\begin{equation}
 \overline D_Q(0)=\overline N_Q(0)=1.
\end{equation}
For $\tau>0$, define
\begin{equation}
 \begin{aligned}
 Q(s;\tau)&:=\overline Q(\tau s)=\frac{N_Q(s;\tau)}{D_Q(s;\tau)},\\
 N_Q(s;\tau)&:=\overline N_Q(\tau s),\qquad
 D_Q(s;\tau):=\overline D_Q(\tau s).
 \end{aligned}
\end{equation}
In particular, $D_Q(0;\tau)=N_Q(0;\tau)=1$.

The conventional inverse-model DOB is specified by the signal equations
\begin{equation}
 y=P(u+d),\qquad
 u=C(r-y)-\widehat d,
 \label{eq:dob_signals_1}
\end{equation}
\begin{equation}
 \widehat d=Q\bigl(P_{\rm n}^{-1}y-u\bigr).
 \label{eq:dob_signals_2}
\end{equation}
Here $r$ is the reference input, $y$ is the measured plant output, $u$ is the control input applied before the additive input disturbance $d$, and $\widehat d$ is the DOB estimate used for input compensation. The estimate incorporates both the input disturbance and the equivalent input effect of plant-model mismatch, since
\[
 P_{\rm n}^{-1}y-u=d+(P_{\rm n}^{-1}P-1)(u+d).
\]
Consequently, the DOB compensates model mismatch in the reference-response analysis with $d=0$.

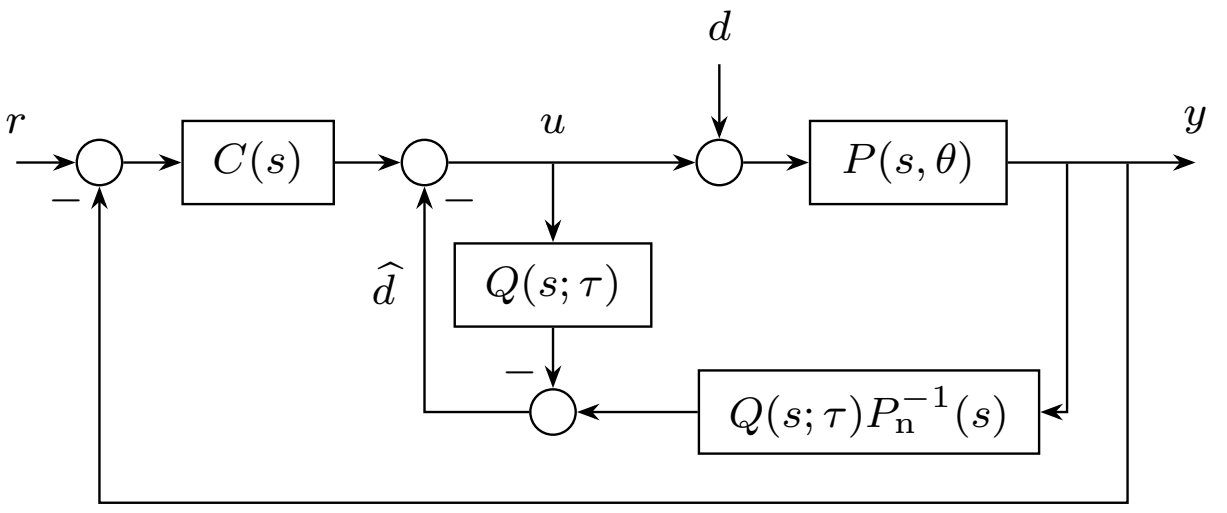
\begin{figure}[H]
\centering
\begin{tikzpicture}[
  x=1cm,y=1cm,
  font=\fontsize{8}{9}\selectfont,
  line width=0.45pt,
  >={Stealth[length=1.5mm,width=1.1mm]},
  block/.style={draw,rectangle,fill=white,minimum height=5.5mm,inner sep=2pt},
  sum/.style={draw,circle,fill=white,minimum size=3mm,inner sep=0pt},
  tap/.style={circle,fill=black,draw=none,inner sep=0pt,minimum size=0.0mm}
]
  \node[sum] (referenceSum) at (0.55,0) {};
  \node[block,minimum width=10mm] (controller) at (1.60,0) {$C(s)$};
  \node[sum] (controlSum) at (2.70,0) {};
  \node[tap] (uTap) at (3.55,0) {};
  \node[sum] (disturbanceSum) at (4.65,0) {};
  \node[block,minimum width=13mm] (plant) at (5.90,0) {$P(s,\theta)$};
  \node[tap] (observerTap) at (6.95,0) {};
  \node[tap] (feedbackTap) at (7.35,0) {};
  \draw[->] (0,0) node[above=2pt] {$r$} -- (referenceSum.west);
  \draw[->] (referenceSum.east) -- (controller.west);
  \draw[->] (controller.east) -- (controlSum.west);
  \draw[->] (controlSum.east) -- (disturbanceSum.west);
  \node[above=2pt] at (uTap) {$u$};
  \draw[->] (4.65,0.65) node[above=1pt] {$d$} -- (disturbanceSum.north);
  \draw[->] (disturbanceSum.east) -- (plant.west);
  \draw[->] (plant.east) -- (7.80,0) node[above=2pt] {$y$};

  \node[block,minimum width=13mm] (qFilter) at (3.55,-0.81) {$Q(s;\tau)$};
  \node[sum] (observerSum) at (3.55,-1.65) {};
  \node[block,minimum width=22.5mm] (filteredInverse) at (5.64,-1.65)
    {$Q(s;\tau)P_{\rm n}^{-1}(s)$};
  \draw[->] (uTap) -- (qFilter.north);
  \draw[->] (qFilter.south) -- (observerSum.north);
  \draw[->] (observerTap) -- (6.95,-1.65) -- (filteredInverse.east);
  \draw[->] (filteredInverse.west) -- (observerSum.east);
  \draw[->] (observerSum.west) -- (2.70,-1.65) --
    node[pos=0.57,left=2pt] {$\widehat d$} (controlSum.south);

  \draw[->] (feedbackTap) -- (7.35,-2.25) -- (0.55,-2.25) -- (referenceSum.south);

  \node at (0.34,0.18) {};
  \node at (0.33,-0.25) {$-$};
  \node at (2.49,0.18) {};
  \node at (2.93,-0.25) {$-$};
  \node at (4.44,0.18) {};
  \node at (4.87,0.24) {};
  \node at (3.33,-1.39) {$-$};
  \node at (3.79,-1.42) {};
\end{tikzpicture}
\caption{Conventional DOB structure.}
\label{fig:dob_structure}
\end{figure}

The signal paths in Fig.~\ref{fig:dob_structure} implement \eqref{eq:dob_signals_1}--\eqref{eq:dob_signals_2}. The output-side observer path realizes the filtered nominal inverse $QP_{\rm n}^{-1}$ as a single block.

A polynomial is called Hurwitz when all its roots lie in the open left half-plane. Internal stability means that, after independent exogenous inputs are introduced at the controller, plant-input, measurement, and DOB-estimate summing junctions, every transfer map to every internal signal is stable and proper.

Define the common algebraic-loop denominator
\begin{equation}
 \Lambda(s;\tau,\theta):=P_{\rm n}(1+PC)+Q(P-P_{\rm n}).
\end{equation}
Define
\begin{equation}
 \mathcal M_{\rm DOB}:=
 \begin{bmatrix}
 1&-P&0\\
 C&1&1\\
 -QP_{\rm n}^{-1}&Q&1
 \end{bmatrix}.
\end{equation}
After independent additive injections $\nu_1,\nu_2,\nu_3$ are introduced into the plant-output, controller, and DOB-estimate equations, respectively, the internal variables satisfy
\begin{equation}
 \mathcal M_{\rm DOB}
 \begin{bmatrix}y\\u\\\widehat d\end{bmatrix}
 =
 \begin{bmatrix}Pd\\Cr\\0\end{bmatrix}
 +
 \begin{bmatrix}\nu_1\\\nu_2\\\nu_3\end{bmatrix}.
 \label{eq:internal_signal_matrix}
\end{equation}
Together, the physical inputs $(r,d)$ and the independent injections $(\nu_1,\nu_2,\nu_3)$ provide independent excitations at the controller, plant-input, measurement, and DOB-estimate junctions.
A direct determinant calculation gives
\begin{equation}
 \det \mathcal M_{\rm DOB}=\frac{\Lambda}{P_{\rm n}}.
 \label{eq:internal_matrix_det}
\end{equation}
Eliminating $\widehat d$ from \eqref{eq:dob_signals_1}--\eqref{eq:dob_signals_2} gives
\begin{equation}
 \begin{bmatrix}y\\u\end{bmatrix}
 =\frac{1}{\Lambda}
 \begin{bmatrix}
 PP_{\rm n}C & PP_{\rm n}(1-Q)\\
 P_{\rm n}C & -P(P_{\rm n}C+Q)
 \end{bmatrix}
 \begin{bmatrix}r\\d\end{bmatrix}.
 \label{eq:yu_transfer_matrix}
\end{equation}
Since $\widehat d=C(r-y)-u$, the maps from $(r,d)$ to $(y,u,\widehat d)$ have $\Lambda$ as their common pre-cancellation closed-loop denominator.

Define the nominal feedback characteristic polynomial
\begin{equation}
 \chi_{\rm n}(s):=D_{\rm n}(s)D_C(s)+N_{\rm n}(s)N_C(s).
\end{equation}
Clearing the chosen coprime block denominators in $\Lambda$ yields the DOB characteristic polynomial
\begin{align}
 \delta(s;\tau,\theta)
 ={}&\bigl(DD_C+NN_C\bigr)N_{\rm n}D_Q \notag\\
 &+N_QD_C\bigl(ND_{\rm n}-N_{\rm n}D\bigr),
 \label{eq:delta_original}
\end{align}
where the arguments of $N,D,N_Q,D_Q$ are suppressed when unambiguous.

Define
\begin{equation}
 \mathcal R_{\rm stab}:=
 \{\tau>0:\delta(\cdot;\tau,\theta)\text{ is Hurwitz for every }\theta\in\Theta\}.
\end{equation}

\begin{assumption}[Standing conditions]
\label{ass:standing}
The following conditions hold.
\begin{enumerate}[label=(A\arabic*),leftmargin=7mm]
\item The nominal interconnection of $P_{\rm n}$ and $C$ is internally stable.
\item The uncertain numerator and denominator degrees are fixed on $\Theta$. The polynomial $\delta(\cdot;\tau,\theta)$ has a common degree $n_\delta$ and a positive leading coefficient on $(0,\infty)\times\Theta$.
\item Fix proper finite-dimensional realizations of $P$, $P_{\rm n}$, $C$, $Q$, and $QP_{\rm n}^{-1}$ for which \eqref{eq:dob_signals_1}--\eqref{eq:dob_signals_2} is well posed. Moreover, $Q(\cdot;\tau)$ and $Q(\cdot;\tau)P_{\rm n}^{-1}$ are stable for every $\tau>0$. Every additional realization denominator is Hurwitz for all $\tau>0$ and $\theta\in\Theta$. In this complete realization, no closed-right-half-plane root of $\delta$ or of an additional realization denominator is cancelled from every transfer map from $(r,d,\nu_1,\nu_2,\nu_3)$ to $(y,u,\widehat d)$ and to the realization states.
\item All polynomial coefficients and uncertainty-box endpoints used in the algebraic computation are rational or real algebraic.
\item The set $\mathcal R_{\rm stab}$ is nonempty.
\end{enumerate}
\end{assumption}

\begin{proposition}[Characteristic polynomial and internal stability]
Under Assumption~\ref{ass:standing}, the interconnection \eqref{eq:dob_signals_1}--\eqref{eq:dob_signals_2} is internally stable for a fixed $(\tau,\theta)$ if and only if $\delta(\cdot;\tau,\theta)$ is Hurwitz.
\end{proposition}

\begin{proof}
Equations~\eqref{eq:internal_signal_matrix}--\eqref{eq:internal_matrix_det} show that every algebraic-loop map from an independent injection to $(y,u,\widehat d)$ is an entry of $\mathcal M_{\rm DOB}^{-1}=\operatorname{adj}(\mathcal M_{\rm DOB})/\det\mathcal M_{\rm DOB}$, where $\operatorname{adj}$ denotes the adjugate matrix. Since $\det\mathcal M_{\rm DOB}=\Lambda/P_{\rm n}$, the factor $\Lambda$ is the common pre-cancellation characteristic factor of these maps. Equation~\eqref{eq:yu_transfer_matrix} gives the same pre-cancellation factor for the physical inputs $(r,d)$. The states of the $Q$-filter and filtered nominal-inverse realizations contribute additional realization denominators that may depend on $\tau$ but are Hurwitz for every $\tau>0$ by Assumption~\ref{ass:standing}(A3). Clearing the coprime denominators of $P$, $P_{\rm n}$, $C$, and $Q$ turns the numerator of $\Lambda$ into $\delta$ in \eqref{eq:delta_original}. Assumption~\ref{ass:standing}(A1) makes the nominal-feedback factors stable, Assumption~\ref{ass:standing}(A2) fixes the degree and sign of the leading coefficient of $\delta$, and Assumption~\ref{ass:standing}(A3) ensures that a closed-right-half-plane root cannot be cancelled from all independently excited internal maps. Hence every internal map is stable and proper exactly when $\delta$ is Hurwitz.
\end{proof}

\begin{proposition}[Constructible robust-stability domain]
\label{prop:Rstab_semialgebraic}
Under Assumption~\ref{ass:standing}, $\mathcal R_{\rm stab}$ is an open semialgebraic subset of $(0,\infty)$. A quantifier-free description of this set can be constructed from the Routh--Hurwitz inequalities by real quantifier elimination.
\end{proposition}

\begin{proof}
The common degree and positive leading coefficient in Assumption~\ref{ass:standing}(A2) make every leading principal Hurwitz determinant of $\delta(\cdot;\tau,\theta)$ a polynomial in $(\tau,\theta)$ with real-algebraic coefficients. Robust stability is therefore described by $\tau>0$, the box inequalities defining $\Theta$, and universal positivity of finitely many polynomial determinants. Tarski--Seidenberg elimination removes $\theta$ and gives a quantifier-free formula in $\tau$. Compactness of $\Theta$ and continuous dependence of the polynomial coefficients make robust Hurwitz stability persistent under sufficiently small changes of $\tau$, so the resulting set is open.
\end{proof}

Consequently, $\mathcal R_{\rm stab}$ is the full constructible domain of robust internal stability and may consist of several disjoint open intervals.

\subsection{Reference-response recovery channel}

For $d=0$, the actual reference-to-output transfer function obtained from \eqref{eq:dob_signals_1}--\eqref{eq:dob_signals_2} is
\begin{equation}
 T_{yr}(s;\tau,\theta)=
 \frac{N_{\rm n}N N_C D_Q}{\delta(s;\tau,\theta)},
 \label{eq:Tyr}
\end{equation}
and the nominal closed-loop reference response is
\begin{equation}
 T_{\rm n}(s)=\frac{N_{\rm n}N_C}{\chi_{\rm n}(s)}.
 \label{eq:Tnom}
\end{equation}
Define
\begin{equation}
 E_r(s;\tau,\theta):=T_{yr}(s;\tau,\theta)-T_{\rm n}(s)
\end{equation}
and the plant-model mismatch polynomial
\begin{equation}
 \Delta_P(s,\theta):=N(s,\theta)D_{\rm n}(s)-N_{\rm n}(s)D(s,\theta).
 \label{eq:DeltaP}
\end{equation}
A rearrangement of \eqref{eq:delta_original} gives
\begin{equation}
 \delta=ND_Q\chi_{\rm n}+D_C(N_Q-D_Q)\Delta_P.
 \label{eq:delta_rearranged}
\end{equation}

\begin{proposition}[Recovery-error factorization]
For every $(\tau,\theta)\in\mathcal R_{\rm stab}\times\Theta$,
\begin{equation}
 E_r(s;\tau,\theta)=
 \frac{N_{\rm n}N_CD_C(D_Q-N_Q)\Delta_P}
 {\chi_{\rm n}\,\delta(s;\tau,\theta)}.
\end{equation}
\end{proposition}

\begin{proof}
Subtracting \eqref{eq:Tnom} from \eqref{eq:Tyr} gives
\[
 E_r=\frac{N_{\rm n}N_C}{\chi_{\rm n}\delta}
 \bigl(ND_Q\chi_{\rm n}-\delta\bigr).
\]
Equation \eqref{eq:delta_rearranged} yields
$ND_Q\chi_{\rm n}-\delta=D_C(D_Q-N_Q)\Delta_P$.
\end{proof}

The factorization separates the plant-model mismatch, the DOB recovery factor $D_Q-N_Q$, and the nominal and uncertain closed-loop dynamics. It also gives
\begin{equation}
 E_r(0;\tau,\theta)=0
\end{equation}
for every robustly stable $\tau$, because $D_Q(0;\tau)=N_Q(0;\tau)$.

Let $W(s)=N_W(s)/D_W(s)$ be a fixed stable real-rational weight represented by a real coprime polynomial pair, with $D_W$ Hurwitz and normalized to have positive leading coefficient, and define
\begin{equation}
 Z(s;\tau,\theta):=W(s)E_r(s;\tau,\theta)
 =\frac{B(s;\tau,\theta)}{A(s;\tau,\theta)},
\end{equation}
where
\begin{align}
 B&:=N_WN_{\rm n}N_CD_C(D_Q-N_Q)\Delta_P,\label{eq:Bdef}\\
 A&:=D_W\chi_{\rm n}\delta.
\end{align}
Let $\mathrm j^2=-1$ and, for a stable proper scalar transfer function $G$, define
\begin{equation}
 \|G\|_{\infty}:=\sup_{\omega\in\mathbb R}|G(\mathrm j\omega)|.
\end{equation}

To interpret the recovery channel, let $y_{\rm n}$ be the nominal output generated by the same reference $r$, with $d=0$ and zero initial conditions. Let $e_W$ be the signal obtained by filtering $y-y_{\rm n}$ through $W$. The transfer function from $r$ to $y-y_{\rm n}$ is $E_r$, and that from $r$ to $e_W$ is $Z$. The weight specifies the frequency emphasis of the criterion: frequencies at which $|W(\mathrm j\omega)|$ is larger receive greater emphasis. With $W=1$, the criterion measures the unweighted departure from the nominal closed-loop response. The complex response difference accounts for both amplitude and phase mismatch.

\begin{assumption}[Performance-channel regularity]
\label{ass:performance}
For every $(\tau,\theta)\in\mathcal R_{\rm stab}\times\Theta$, the polynomial $A(\cdot;\tau,\theta)$ is Hurwitz, has a common degree $n_A$, and has a nonzero leading coefficient. Moreover,
\begin{equation}
 \deg_s B(s;\tau,\theta)<n_A.
\end{equation}
\end{assumption}

The nominal model, nominal controller, normalized $Q$-filter shape, performance weight, and uncertainty set are prescribed. The design variable is the single time constant $\tau$, which scales the bandwidth of the prescribed $Q$-filter. The constraint $\tau\in\mathcal R_{\rm stab}$ ensures internal stability for every admissible plant. The objective evaluates the largest weighted response mismatch over the uncertainty set and all frequencies. The squared robust recovery value and its possibly empty optimizer set are
\begin{equation}
 J(\tau):=
 \sup_{\theta\in\Theta}\left\|Z(\cdot;\tau,\theta)\right\|_{\infty}^{2},
 \qquad \tau\in\mathcal R_{\rm stab},
 \label{eq:robust_recovery_value}
\end{equation}
\begin{equation}
 \mathcal T^{\star}:=
 \operatorname*{arg\,min}_{\tau\in\mathcal R_{\rm stab}}J(\tau).
 \label{eq:optimal_time_constants}
\end{equation}
Let $\|\cdot\|_{L_2}$ denote the norm of a square-integrable time signal on $[0,\infty)$. With zero initial conditions and $d=0$, every square-integrable reference satisfies
\[
 \|e_W\|_{L_2}\le\sqrt{J(\tau)}\,\|r\|_{L_2}
 \qquad\text{for every fixed }\theta\in\Theta.
\]
Thus $\sqrt{J(\tau)}$ is the worst-case induced $L_2$ gain from the reference to the weighted nominal-response mismatch. Squaring the norm preserves the minimizing time constants in \eqref{eq:intro_problem} and facilitates polynomial performance inequalities. The set $\mathcal T^{\star}$ contains all time constants that attain the smallest worst-case mismatch over the complete robust-stability domain. Because this domain can contain several open intervals, the formulation accounts for all components and explicitly determines whether the infimum is attained. When $\mathcal T^{\star}$ is a singleton, its element is denoted by $\tau^{\star}$.

High-bandwidth DOB analysis explains nominal-performance recovery under its stability assumptions \citep{Shim2016}. The present criterion evaluates finite time constants through both the recovery factor and the uncertain closed-loop dynamics. Closed-loop $H_{\infty}$ design with a prescribed outer-loop controller is also established in DOB design: \citet{WangTomizuka2004} optimize observer gains through a static-output-feedback formulation, and nonsmooth methods support broader fixed-order and structured parameterizations through numerical optimization \citep{ApkarianNoll2006,GumussoyOverton2008}. For the prescribed scalar family $Q(s;\tau)=\overline Q(\tau s)$, the task here is to determine the global infimum, its attainment status, and every minimizing time constant over the full robust-stability domain.

\section{Algebraic reduction of the robust $H_{\infty}$ inequality}

\subsection{Shared affine uncertainty}

Let $\mathbb K\subset\mathbb R$ denote the field generated by the real-algebraic coefficients of the plant, nominal model, controller, normalized $Q$-filter, performance weight, and uncertainty-box endpoints.

\begin{lemma}[Affine numerator-denominator dependence]
There exist polynomial families $A_k^{\rm aff}(s;\tau)$ and $B_k^{\rm aff}(s;\tau)$ such that
\begin{align}
 A(s;\tau,\theta)&=A_0^{\rm aff}(s;\tau)+\sum_{k=1}^{p}\theta_kA_k^{\rm aff}(s;\tau),\\
 B(s;\tau,\theta)&=B_0^{\rm aff}(s;\tau)+\sum_{k=1}^{p}\theta_kB_k^{\rm aff}(s;\tau).
\end{align}
\end{lemma}

\begin{proof}
The characteristic polynomial \eqref{eq:delta_original} is affine in $N$ and $D$, hence in $\theta$, so $A=D_W\chi_{\rm n}\delta$ is affine. The mismatch polynomial \eqref{eq:DeltaP} is affine in the same vector, and all remaining factors in \eqref{eq:Bdef} are fixed for a given $\tau$.
\end{proof}

For a prescribed level $\gamma>0$, directly forming $\gamma^2|A(\mathrm j\omega)|^2-|B(\mathrm j\omega)|^2$ makes the uncertainty dependence quadratic. Phase lifting preserves the shared affine representation before magnitude squares are taken.

\subsection{Phase lifting and polynomial edges}

\begin{lemma}[Phase-lifting equivalence]
\label{lem:phase_lifting}
Let $A$ be Hurwitz and let $B/A$ be strictly proper. For every $\gamma>0$,
\begin{equation}
\begin{aligned}
 \left\|\frac{B}{A}\right\|_{\infty}<\gamma
 \quad\Longleftrightarrow\quad&
 A(s)+\gamma^{-1}e^{\mathrm j\phi}B(s)\text{ is Hurwitz}\\
 &\text{for every }\phi\in[0,2\pi).
\end{aligned}
\end{equation}
\end{lemma}

\begin{proof}
If the norm is below $\gamma$, the homotopy
$A+t\gamma^{-1}e^{\mathrm j\phi}B$, $t\in[0,1]$, cannot acquire an imaginary-axis root, because such a root would imply $|B/A|=\gamma/t\ge\gamma$. Strict properness preserves the degree and leading coefficient throughout the homotopy, preventing a root crossing at infinity. Conversely, if the norm is at least $\gamma$, strict properness and continuity give a finite frequency at which $|B/A|=\gamma$; an appropriate phase then places a zero of the lifted polynomial on the imaginary axis.
\end{proof}

For fixed $(\tau,\gamma,\phi)$, define the lifted polynomial
\begin{equation}
 p_{\phi}(s;\tau,\theta)
 :=A(s;\tau,\theta)+\gamma^{-1}e^{\mathrm j\phi}B(s;\tau,\theta).
\end{equation}
Evaluate the pair $(A,B)$ at every vertex of $\Theta$ and retain one representative of each distinct pair in $\mathbb K[s,\tau]^2$. Let $m$ be the number of retained pairs and let $v^{(1)},\ldots,v^{(m)}\in\Theta$ be their representative vertices. Define
\begin{equation}
 A_i(s;\tau):=A(s;\tau,v^{(i)}),\qquad
 B_i(s;\tau):=B(s;\tau,v^{(i)}),
\end{equation}
and
\begin{equation}
 p_{\phi,i}(s;\tau):=A_i(s;\tau)+\gamma^{-1}e^{\mathrm j\phi}B_i(s;\tau).
\end{equation}
Let $\operatorname{co}$ denote the convex hull. The shared affine dependence gives the exact convex-hull identity
\begin{equation}
 \{p_{\phi}(\cdot;\tau,\theta):\theta\in\Theta\}
 =\operatorname{co}\{p_{\phi,1},\ldots,p_{\phi,m}\}.
\end{equation}
For $i<j$ and $\lambda\in[0,1]$, set
\begin{align}
 A_{ij}(s;\tau,\lambda)&:=\lambda A_i(s;\tau)+(1-\lambda)A_j(s;\tau),\\
 B_{ij}(s;\tau,\lambda)&:=\lambda B_i(s;\tau)+(1-\lambda)B_j(s;\tau).
\end{align}

\begin{theorem}[Non-strict robust-performance edge reduction]
\label{thm:edge_reduction}
Fix $\tau\in\mathcal R_{\rm stab}$ and $\gamma>0$. Under Assumptions~\ref{ass:standing} and~\ref{ass:performance},
\begin{equation}
 \sup_{\theta\in\Theta}
 \left\|\frac{B(\cdot;\tau,\theta)}{A(\cdot;\tau,\theta)}\right\|_{\infty}
 \le\gamma
 \label{eq:full_closed}
\end{equation}
if and only if
\begin{equation}
 \left\|\frac{B_{ij}(\cdot;\tau,\lambda)}{A_{ij}(\cdot;\tau,\lambda)}\right\|_{\infty}
 \le\gamma
 \label{eq:edge_closed}
\end{equation}
for every pair $i<j$ and every $\lambda\in[0,1]$. If $m=1$, only the single corner condition is required.
\end{theorem}

\begin{proof}
For every pair and every $\lambda\in[0,1]$, the point $v_{ij}(\lambda):=\lambda v^{(i)}+(1-\lambda)v^{(j)}$ belongs to the box $\Theta$, and shared affine dependence gives $(A_{ij},B_{ij})=(A,B)(\cdot;\tau,v_{ij}(\lambda))$. Thus every tested segment belongs to the physical uncertainty image, proving necessity. For sufficiency, fix $\varepsilon>0$ and $\phi$, and define the lifted family at level $\gamma+\varepsilon$ by
\begin{equation}
 p_{\phi,\varepsilon}(s;\tau,\theta)
 :=A(s;\tau,\theta)+(\gamma+\varepsilon)^{-1}e^{\mathrm j\phi}B(s;\tau,\theta).
\end{equation}
Equation~\eqref{eq:edge_closed} implies a strict bound $\gamma+\varepsilon$ on every pairwise segment. Lemma~\ref{lem:phase_lifting}, applied at this level, makes every edge of the polytope generated by $p_{\phi,\varepsilon}$ Hurwitz. By Assumption~\ref{ass:performance}, this lifted family has fixed degree, and the phase lift leaves each member's nonzero leading coefficient equal to that of $A$. The fixed-degree complex-polynomial Edge theorem \citep{ShiZhang1995,BialasGora2012} therefore makes the entire lifted polytope Hurwitz. Applying Lemma~\ref{lem:phase_lifting} gives $\|B(\cdot;\tau,\theta)/A(\cdot;\tau,\theta)\|_{\infty}<\gamma+\varepsilon$ for every $\theta\in\Theta$. Letting $\varepsilon\downarrow0$ for each $\theta$ and then taking the supremum proves \eqref{eq:full_closed}.
\end{proof}

The finite all-pairs formulation avoids tracking how the exposed-edge set changes with $(\tau,\gamma,\phi)$; for fixed $(\tau,\gamma,\phi)$, only the exposed edges are theoretically required.

\subsection{Frequency polynomial and analytic edge elimination}

Set
\begin{equation}
 \eta:=\omega^2\ge0,\qquad g:=\gamma^2.
\end{equation}
For a real polynomial $F$, write
\begin{equation}
 F(\mathrm j\omega)=F_R(\eta)+\mathrm j\omega F_I(\eta),
\end{equation}
so that $|F(\mathrm j\omega)|^2=[F_R(\eta)]^2+\eta[F_I(\eta)]^2$ is polynomial in $\eta$.

For each pair $(i,j)$, define
\begin{align}
 H_{ij}(\eta,\tau,\lambda;g):={}&
 g\left|A_{ij}(\mathrm j\sqrt\eta;\tau,\lambda)\right|^2 \notag\\
 &-\left|B_{ij}(\mathrm j\sqrt\eta;\tau,\lambda)\right|^2.
\end{align}
Because $A_{ij}$ and $B_{ij}$ are affine in $\lambda$,
\begin{equation}
 H_{ij}=h_{2,ij}\lambda^2+h_{1,ij}\lambda+h_{0,ij},
 \label{eq:Hquadratic}
\end{equation}
where each coefficient is polynomial in $(\eta,\tau,g)$. Define
\begin{equation}
 \mathscr D_{ij}:=4h_{2,ij}h_{0,ij}-h_{1,ij}^2.
\end{equation}

\begin{lemma}[Nonnegative quadratic on the unit interval]
\label{lem:quadratic_nonnegative}
The quadratic $q(\lambda)=h_2\lambda^2+h_1\lambda+h_0$ is nonnegative for every $\lambda\in[0,1]$ if and only if
\begin{equation}
 h_0\ge0,\qquad h_0+h_1+h_2\ge0,
\end{equation}
and
\begin{equation}
 (h_2\le0)\ \vee\ (h_1\ge0)\ \vee\ (h_1\le-2h_2)\ \vee\ (4h_2h_0-h_1^2\ge0).
\end{equation}
\end{lemma}

\begin{proof}
For $h_2\le0$, the minimum occurs at an endpoint. For $h_2>0$, the unconstrained minimizer is $-h_1/(2h_2)$. It lies at or to the left of $0$ when $h_1\ge0$, at or to the right of $1$ when $h_1\le-2h_2$, and in $(0,1)$ otherwise. In the interior case the minimum is $(4h_2h_0-h_1^2)/(4h_2)$.
\end{proof}

For a corner $i$, define
\begin{equation}
 H_i(\eta,\tau;g):=
 g|A_i(\mathrm j\sqrt\eta;\tau)|^2-|B_i(\mathrm j\sqrt\eta;\tau)|^2.
\end{equation}
The endpoint values of \eqref{eq:Hquadratic} are $H_j$ and $H_i$.

\begin{theorem}[Robust-performance predicate]
\label{thm:closed_test}
For $\tau\in\mathcal R_{\rm stab}$ and $g>0$, the inequality $J(\tau)\le g$ holds if and only if
\begin{enumerate}[label=(\roman*),leftmargin=5mm]
\item $H_i(\eta,\tau;g)\ge0$ for every corner $i$ and every $\eta\ge0$;
\item for every pair $i<j$ and every $\eta\ge0$,
\begin{equation}
\begin{aligned}
 &(h_{2,ij}\le0)\vee(h_{1,ij}\ge0)\\
 &\qquad\vee(h_{1,ij}\le-2h_{2,ij})\vee(\mathscr D_{ij}\ge0).
\end{aligned}
\end{equation}
\end{enumerate}
All quantities are real polynomials in $(\eta,\tau,g)$.
\end{theorem}

\begin{proof}
Theorem~\ref{thm:edge_reduction} reduces the robust norm bound to the pairwise segments. On each segment, the bound is equivalent to $H_{ij}\ge0$ for every $\eta\ge0$ and $\lambda\in[0,1]$. Lemma~\ref{lem:quadratic_nonnegative} eliminates $\lambda$, with the two endpoint inequalities supplied by the corner conditions.
\end{proof}

The quantified dimension is reduced as
\begin{equation}
\begin{aligned}
 \underbrace{\forall\eta\ \forall\theta_1\cdots\forall\theta_p}_{p+1\ \text{real variables}}
 &\longrightarrow
 \underbrace{\forall\eta\ \forall\lambda}_{2\ \text{real variables}}\\
 &\longrightarrow
 \underbrace{\forall\eta}_{1\ \text{real variable}}.
\end{aligned}
 \label{eq:dimension_reduction}
\end{equation}
The first arrow is the polynomial-edge reduction; the second is the analytic quadratic test. For the DOB family considered here, this reduction leaves only the squared-frequency variable for real quantifier elimination.

\section{Finite computation of the global infimum and optimizer set}
\label{sec:optimizer}

\noindent\textbf{Input.}
The computation takes the affine plant-uncertainty box, the nominal model and controller, the prescribed $Q$-filter shape, and the performance weight.

\medskip
\noindent\textbf{Output.}
It returns the global infimum of $J$, an attainment decision, and a semialgebraic representation of the complete optimizer set.

Define the epigraph of $J$ relative to the robust-stability domain by
\begin{equation}
 \operatorname{epi}_{\mathcal R_{\rm stab}}J:=
 \{(\tau,g)\in\mathcal R_{\rm stab}\times[0,\infty):J(\tau)\le g\}.
\end{equation}
The auxiliary variable $g$ polynomializes the $H_{\infty}$ inequality. For $g=0$, membership is equivalent to the polynomial identity
\begin{equation}
 B(s;\tau,\theta)\equiv0
 \quad\text{in }(s,\theta),
\end{equation}
which is decided by coefficient vanishing. For $g>0$, Theorem~\ref{thm:closed_test} supplies the required first-order real-algebraic formula.

\begin{theorem}[Finite computation of the global infimum and optimizer set]
\label{thm:optimizer_computable}
Under Assumptions~\ref{ass:standing} and~\ref{ass:performance}, the following statements hold.
\begin{enumerate}[label=(\roman*),leftmargin=5mm]
\item $\operatorname{epi}_{\mathcal R_{\rm stab}}J$ is semialgebraic and has a finite quantifier-free representation.
\item The value
\begin{equation}
 g^{\star}:=\inf_{\tau\in\mathcal R_{\rm stab}}J(\tau)
\end{equation}
is finite and real algebraic.
\item Whether the infimum is attained is decidable by real quantifier elimination followed by exact root isolation and sign determination.
\item If it is attained, the full optimizer set
\begin{equation}
 \mathcal T^{\star}=
 \{\tau\in\mathcal R_{\rm stab}:(\tau,g^{\star})\in
 \operatorname{epi}_{\mathcal R_{\rm stab}}J\}
 \label{eq:Tstar_epigraph}
\end{equation}
can be constructed as a semialgebraic subset of the real line.
\item Every isolated element of $\mathcal T^{\star}$ is real algebraic and admits an algebraic isolating representation.
\end{enumerate}
\end{theorem}

\begin{proof}
Proposition~\ref{prop:Rstab_semialgebraic}, Theorem~\ref{thm:closed_test}, and the coefficient-vanishing condition at $g=0$ express the relative epigraph by a first-order formula over $\mathbb K$. Tarski--Seidenberg elimination gives a quantifier-free representation. Since $\mathcal R_{\rm stab}$ is nonempty, selecting any $\tau_0\in\mathcal R_{\rm stab}$ gives $0\le g^{\star}\le J(\tau_0)<\infty$; compactness of $\Theta$, robust stability, and strict properness ensure the final inequality. The projection of the epigraph onto the $g$-axis is semialgebraic, so its lower endpoint is real algebraic. The fiber at that endpoint decides attainment and, when nonempty, gives \eqref{eq:Tstar_epigraph}. Standard real-root isolation yields the isolated elements \citep{Basu2006}.
\end{proof}

The proof yields the following finite computation.
\begin{enumerate}[label=\textbf{Step \arabic*:},leftmargin=15mm]
\item Construct $\mathcal R_{\rm stab}$ from the Routh--Hurwitz inequalities by real quantifier elimination.
\item Construct the weighted recovery channel $B/A$ and its shared affine uncertainty representation.
\item Enumerate the distinct joint corner-image polynomial families and the finite polynomial-edge conditions.
\item Eliminate the edge coordinate analytically and form the one-frequency predicate of Theorem~\ref{thm:closed_test}.
\item Apply real quantifier elimination in the squared-frequency variable to obtain the epigraph in $(\tau,g)$.
\item Project the epigraph onto the $g$-axis, isolate its lower endpoint $g^{\star}$, and decide whether the endpoint fiber is nonempty.
\item Construct the complete optimizer set $\mathcal T^{\star}$ and isolate every isolated optimizer algebraically.
\end{enumerate}
In structured cases, algebraic candidates may instead be isolated first and then proved globally optimal.

Suppose a candidate optimum has squared peak frequency $\eta_c>0$, an interior time constant $\tau_c\in\mathcal R_{\rm stab}$, and squared performance level $g_c>0$, with a locally unique smooth active mechanism. Algebraic active-peak equations can then accelerate candidate generation. A smooth corner peak satisfies
\begin{equation}
\begin{aligned}
 H_i(\eta_c,\tau_c;g_c)&=0,\\
 \partial_{\eta}H_i(\eta_c,\tau_c;g_c)&=0,\\
 \partial_{\tau}H_i(\eta_c,\tau_c;g_c)&=0.
\end{aligned}
\end{equation}
A smooth interior-edge peak satisfies
\begin{equation}
\begin{aligned}
 \mathscr D_{ij}(\eta_c,\tau_c;g_c)&=0,\\
 \partial_{\eta}\mathscr D_{ij}(\eta_c,\tau_c;g_c)&=0,\\
 \partial_{\tau}\mathscr D_{ij}(\eta_c,\tau_c;g_c)&=0,
\end{aligned}
\end{equation}
together with $h_{2,ij}>0$ and $-2h_{2,ij}<h_{1,ij}<0$, evaluated at $(\eta_c,\tau_c,g_c)$, as in Lemma~\ref{lem:quadratic_nonnegative}. Minimax optima can also equalize several stationary peaks. These equations generate candidate points, while the epigraph computation checks every robust-stability component and active mechanism.

\section{Case study: global time-constant tuning with NMP plant uncertainty}

This case study applies the preceding framework to a one-parameter uncertain plant family containing both minimum-phase and non-minimum-phase members. The nominal controller and normalized $Q$-filter shape are fixed, so the design variable is the filter time constant. The calculation determines the complete robust-stability domain and certifies the global minimizer of the worst-case nominal-response-recovery error. For this family, candidate isolation and a direct global-exclusion argument provide the certificate without constructing the full performance epigraph.

\subsection{Problem data and robust-stability domain}

Consider
\begin{equation}
 C(s)=\frac{1}{s+1},\qquad
 Q(s;\tau)=\frac{3\tau s+1}{\tau^2s^2+8\tau s+1},
 \label{eq:example_CQ}
\end{equation}
\begin{align}
 P_{\beta}(s)&=\frac{s^2+\beta s+5}{(s+1)^3},
 &\beta&\in\left[-\frac15,2\right],\notag\\
 P_{\rm n}(s)&=\frac{s^2+s+5}{(s+1)^3}.
 \label{eq:example_P}
\end{align}
At $\beta=-1/5$, the numerator roots have real part $1/10>0$, whereas the nominal model corresponds to $\beta=1$ and is minimum phase. Define $\delta_{\beta}(s;\tau):=\delta(s;\tau,\beta)$. Direct substitution into \eqref{eq:delta_original} gives
\begingroup\small
\begin{align}
\delta_{\beta}(s;\tau)={}&\tau^2s^8+(5\tau^2+8\tau)s^7\notag\\
&+(3\beta\tau+16\tau^2+37\tau+1)s^6\notag\\
&+(\beta\tau^2+12\beta\tau+\beta+31\tau^2
 +116\tau+4)s^5\notag\\
&+(\beta\tau^2+26\beta\tau+4\beta+45\tau^2
 +230\tau+12)s^4\notag\\
&+(5\beta\tau^2+20\beta\tau+7\beta+26\tau^2
 +348\tau+25)s^3\notag\\
&+(43\beta\tau+5\beta+30\tau^2\notag\\
&\qquad +205\tau+41)s^2\notag\\
&+(6\beta+240\tau+25)s+30.
\end{align}
\endgroup
The polynomial has degree eight and positive leading coefficient $\tau^2$ for every $\tau>0$. For a degree-$n$ polynomial
\[
 \varpi(s)=\alpha_0s^n+\alpha_1s^{n-1}+\cdots+\alpha_n,
\]
let $\mathcal H(\varpi)$ be the Hurwitz matrix whose first rows are
\[
 \begin{bmatrix}
 \alpha_1&\alpha_3&\alpha_5&\cdots\\
 \alpha_0&\alpha_2&\alpha_4&\cdots\\
 0&\alpha_1&\alpha_3&\cdots\\
 0&\alpha_0&\alpha_2&\cdots\\
 \vdots&\vdots&\vdots&\ddots
 \end{bmatrix},
\]
and let $\Delta_k^{\rm H}(\varpi)$ be the determinant of its leading $k\times k$ principal submatrix. For brevity, define
\begin{equation}
 M_k(\tau,\beta):=\Delta_k^{\rm H}\!\left(\delta_\beta(\cdot;\tau)\right),
 \qquad k=1,\ldots,8.
\end{equation}

The robust-stability formula is
\begin{equation}
\begin{aligned}
 &\tau>0,\\
 &M_k(\tau,\beta)>0
 \quad \forall\beta\in[-1/5,2],\quad k=1,\ldots,8.
\end{aligned}
 \label{eq:case_stability_formula}
\end{equation}
Here $\operatorname{lc}_{\beta}$ and $\operatorname{Res}_{\beta}$ denote
the leading coefficient and resultant in $\beta$. Let $\mathscr P_{\rm stab}$
be the set of distinct nonconstant
primitive irreducible factors in $\mathbb Q[\tau]$ of the nonzero
polynomials among $\operatorname{lc}_{\beta}M_k$,
$M_k(\tau,-1/5)$, $M_k(\tau,2)$, and
$\operatorname{Res}_{\beta}(M_k,\partial_{\beta}M_k)$ for those
$k$ with $\deg_{\beta}M_k\ge1$, $k=1,\ldots,8$. Zero and constant
projection polynomials are discarded before factorization. Between
consecutive positive roots of this set, each $M_k(\tau,\cdot)$ has
fixed degree and no multiple or endpoint-crossing root, so
\eqref{eq:case_stability_formula} has a constant truth value. The
set contains $25$ factors with $29$ distinct positive roots. One
factor is
\begingroup\scriptsize
\begin{align}
P_{12}(\tau)={}&513812250000\tau^{12}
 +11696457828750\tau^{11}\notag\\
&+36327228201925\tau^{10}
 +71724533500640\tau^9\notag\\
&-8770348113652\tau^8
 -193829557926750\tau^7\notag\\
&-917229606840206\tau^6
 +660614449088377\tau^5\notag\\
&-67380186556048\tau^4
 -7532171680402\tau^3\notag\\
&-132726255603\tau^2
 +8488218829\tau+240943180.
\end{align}
\endgroup
In particular,
\begin{equation}
 M_7(\tau,-1/5)=-\frac{2\tau}{15625}P_{12}(\tau),
 \qquad \gcd(P_{12},P_{12}')=1.
\end{equation}

For $\sigma>0$ and $a\le b$, the notation $\sigma[a,b]$ means the interval $[\sigma a,\sigma b]$.

\begin{proposition}[Exact robust-stability decomposition]
The polynomial $P_{12}$ has exactly four positive roots
\[
 0<\tau_1^{\rm s}<\tau_2^{\rm s}<\tau_3^{\rm s}<\tau_4^{\rm s},
\]
isolated by
\begin{align}
\tau_1^{\rm s}&\in10^{-14}[3455203832122,3455203832123],\notag\\
\tau_2^{\rm s}&\in10^{-14}[23213330934976,23213330934977],\notag\\
\tau_3^{\rm s}&\in10^{-14}[48857860637616,48857860637617],\notag\\
\tau_4^{\rm s}&\in10^{-14}[174838134330976,174838134330977].
\label{eq:stability_endpoint_boxes}
\end{align}
Moreover,
\begin{equation}
 \mathcal R_{\rm stab}
 = (\tau_1^{\rm s},\tau_2^{\rm s})
 \cup(\tau_3^{\rm s},\tau_4^{\rm s})
 =\{\tau>0:P_{12}(\tau)<0\}.
 \label{eq:example_Rs_exact}
\end{equation}
\end{proposition}

\begin{proof}
Exact evaluation of all $29$ singleton root cells and their
adjacent open cells shows that only the four roots of $P_{12}$ change
the truth value of \eqref{eq:case_stability_formula}. At those roots
$M_7(\tau,-1/5)=0$; at every other positive projection root the
singleton value equals the common adjacent-cell value. Hence, on
$\tau>0$, \eqref{eq:case_stability_formula} is equivalent to
$P_{12}(\tau)<0$. Let $V(a)$ be the sign-variation count of the
Sturm sequence of $P_{12}$ at $a$, with zero entries omitted; at
$+\infty$, use the leading coefficients of the Sturm polynomials.
Exact rational evaluation gives $V(0)=7$, $V(+\infty)=3$, and
\begin{center}
\scriptsize
\begin{tabular}{@{}ccc@{}}
\toprule
Root interval & $V(\mathrm{left})$ & $V(\mathrm{right})$\\
\midrule
interval for $\tau_1^{\rm s}$ & $7$ & $6$\\
interval for $\tau_2^{\rm s}$ & $6$ & $5$\\
interval for $\tau_3^{\rm s}$ & $5$ & $4$\\
interval for $\tau_4^{\rm s}$ & $4$ & $3$\\
\bottomrule
\end{tabular}
\end{center}
Sturm's theorem therefore gives exactly one root in each displayed interval and no other positive root. On the five resulting open cells, the truth value of
\eqref{eq:case_stability_formula} is constant. For
$\Pi(\varrho)=\sum_{\mu=0}^{n_{\Pi}}c_{\mu}\varrho^{\mu}$ and
$n_{\rm B}\ge n_{\Pi}$, define its degree-$n_{\rm B}$ Bernstein
coefficients by
\begin{equation}
b_{\iota}^{(n_{\rm B})}
=\sum_{\mu=0}^{\min(\iota,n_{\Pi})}c_{\mu}
\frac{\binom{\iota}{\mu}}{\binom{n_{\rm B}}{\mu}},
\qquad \iota=0,\ldots,n_{\rm B}.
\label{eq:bernstein_coefficients}
\end{equation}
Define $\beta(\xi):=-1/5+(11/5)\xi$ for $\xi\in[0,1]$.
For a rational sample point $\tau_{\rm samp}$ in a given open cell, let
$n_k:=\deg_{\xi}M_k(\tau_{\rm samp},\beta(\xi))$, denote its
natural-degree Bernstein coefficients by
$b_{k,\iota}^{(n_k)}(\tau_{\rm samp})$, and set
$b_{\min}(\tau_{\rm samp}):=
\min_{1\le k\le8,\,0\le\iota\le n_k}
 b_{k,\iota}^{(n_k)}(\tau_{\rm samp})$. The selected sample points and
exact sign tests are
\begin{center}
\scriptsize
\begin{tabular}{@{}cccc@{}}
\toprule
Cell & $\tau_{\rm samp}$ & Sign test & Class\\
\midrule
$(0,\tau_1^{\rm s})$ & $1/100$ & $M_7(\tau_{\rm samp},-1/5)<0$ & unstable\\
$(\tau_1^{\rm s},\tau_2^{\rm s})$ & $1/10$ & $b_{\min}(\tau_{\rm samp})\ge17/20$ & stable\\
$(\tau_2^{\rm s},\tau_3^{\rm s})$ & $3/10$ & $M_7(\tau_{\rm samp},-1/5)<0$ & unstable\\
$(\tau_3^{\rm s},\tau_4^{\rm s})$ & $1$ & $b_{\min}(\tau_{\rm samp})\ge13$ & stable\\
$(\tau_4^{\rm s},\infty)$ & $2$ & $M_7(\tau_{\rm samp},-1/5)<0$ & unstable\\
\bottomrule
\end{tabular}
\end{center}
This proves \eqref{eq:example_Rs_exact}.
\end{proof}

For interpretation, the two components are approximately
\begin{equation}
 (0.034552,0.232133)\cup(0.488579,1.748381).
\end{equation}
For this family,
\begin{equation}
 \chi_{\rm n}(s)=s^4+4s^3+7s^2+5s+6,
\end{equation}
and
\begin{equation}
 \Delta_P(s,\beta)=(\beta-1)s(s+1)^3,\qquad
 D_Q-N_Q=\tau s(\tau s+5).
\end{equation}
With $W=1$, define
\begin{align}
 \mathcal N_{\beta}(s,\tau)
 &:=(\beta-1)\tau s^2(\tau s+5)(s+1)^4(s^2+s+5),\\
 \mathcal A_{\beta}(s,\tau)
 &:=\chi_{\rm n}(s)\delta_{\beta}(s;\tau),
\end{align}
so that
\begin{equation}
 E_{\beta}(s;\tau):=E_r(s;\tau,\beta)=\frac{\mathcal N_{\beta}(s,\tau)}{\mathcal A_{\beta}(s,\tau)}.
\end{equation}
The squared minimax objective and its optimizer set are
\begin{align}
 J_{\rm ex}(\tau)&:=
 \sup_{\beta\in[-1/5,2]}\|E_{\beta}(\cdot;\tau)\|_{\infty}^{2},
 \\
 \mathcal T_{\rm ex}^{\star}&:=
 \operatorname*{arg\,min}_{\tau\in\mathcal R_{\rm stab}}J_{\rm ex}(\tau).
\end{align}

Using $\eta=\omega^2$, define
\begin{align}
 N_{\beta}(\eta,\tau)&:=|\mathcal N_{\beta}(\mathrm j\sqrt\eta,\tau)|^2,\\
 D_{\beta}(\eta,\tau)&:=|\mathcal A_{\beta}(\mathrm j\sqrt\eta,\tau)|^2,\\
 H_{\beta}(\eta,\tau,g)&:=gD_{\beta}(\eta,\tau)-N_{\beta}(\eta,\tau).
\end{align}
For $\tau\in\mathcal R_{\rm stab}$,
\begin{equation}
\begin{aligned}
 J_{\rm ex}(\tau)\le g
 \quad\Longleftrightarrow\quad&
 H_{\beta}(\eta,\tau,g)\ge0,\\[-1mm]
 &\forall\eta\ge0,\quad \forall\beta\in[-1/5,2].
\end{aligned}
\end{equation}
Define the endpoint restrictions $H_-:=H_{\beta}|_{\beta=-1/5}$ and $H_+:=H_{\beta}|_{\beta=2}$. The endpoint two-peak system at $\beta=-1/5$ generates the algebraic candidate, while the global argument covers the complete robust-stability domain.

\subsection{Algebraic candidate isolation}

Introduce two distinct candidate squared frequencies $\eta_1,\eta_2>0$. The equal-peak stationarity conditions are
\begin{equation}
\begin{aligned}
 H_-(\eta_1,\tau,g)&=0,&\qquad \partial_{\eta}H_-(\eta_1,\tau,g)&=0,\\
 H_-(\eta_2,\tau,g)&=0,& \partial_{\eta}H_-(\eta_2,\tau,g)&=0.
\end{aligned}
\label{eq:equal_peak_system}
\end{equation}
The following rational box serves as an isolating witness for the algebraic candidate:
\begin{equation}
 \mathcal X:=I_{\eta,1}\times I_{\eta,2}\times I_{\tau}\times I_g,
\end{equation}
where
\begingroup\scriptsize
\begin{align}
 I_{\eta,1}&:=10^{-32}[114078503953394758508281580701662,\notag\\
 &\hspace{13mm}114078503953394758508281580701663],\\
 I_{\eta,2}&:=10^{-32}[425278838028374088289400499869045,\notag\\
 &\hspace{13mm}425278838028374088289400499869047],\\
 I_{\tau}&:=10^{-32}[3640745160795200590078581763342,\notag\\
 &\hspace{13mm}3640745160795200590078581763343],\label{eq:Itau_exact}\\
 I_g&:=10^{-32}[528885058591472717658155303878118,\notag\\
 &\hspace{13mm}528885058591472717658155303878119].
\end{align}
\endgroup
The first two intervals are disjoint. Let $\boldsymbol{x}:=(\eta_1,\eta_2,\tau,g)$ and define
\[
\resizebox{0.98\columnwidth}{!}{$
\boldsymbol F(\boldsymbol{x}):=\bigl(H_-(\eta_1,\tau,g),\partial_\eta H_-(\eta_1,\tau,g),H_-(\eta_2,\tau,g),\partial_\eta H_-(\eta_2,\tau,g)\bigr)^{\mathsf T}
$}.
\]
\begin{proposition}[Algebraic optimizer candidate]
The system \eqref{eq:equal_peak_system} has exactly one real solution in $\mathcal X$. Denote it by
\begin{equation}
 (\widehat\eta_1,\widehat\eta_2,\widehat\tau,\widehat g).
\end{equation}
Its numerical values relevant to the design are
\begin{equation}
 \widehat\tau\approx0.0364075,\qquad
 \sqrt{\widehat g}\approx2.29975,
\end{equation}
with candidate stationary frequencies
\begin{equation}
 \sqrt{\widehat\eta_1}\approx1.06808,\qquad
 \sqrt{\widehat\eta_2}\approx2.06223.
 \label{eq:candidate_frequency_values}
\end{equation}
Every coordinate is real algebraic.
\end{proposition}

\begin{proof}
Let $[D\boldsymbol F](\mathcal X)$ be the natural rational interval enclosure of the Jacobian of $\boldsymbol F$ on $\mathcal X$, and let $I_4$ be the $4\times4$ identity matrix.
Let $\boldsymbol{x}_0$ be the componentwise midpoint of $\mathcal X$. Define $Y\in\mathbb Q^{4\times4}$ by
\begin{equation}
\resizebox{0.98\columnwidth}{!}{$
Y:=\begin{bmatrix}
-6.118\!\times\!10^{-4} & 1.8396\!\times\!10^{-4} & 1.496\!\times\!10^{-7} & 0\\
4.2631\!\times\!10^{-4} & 0 & -2.9717\!\times\!10^{-6} & 1.3429\!\times\!10^{-7}\\
-4.3403\!\times\!10^{-5} & 0 & 2.8711\!\times\!10^{-7} & 0\\
2.3612\!\times\!10^{-1} & 0 & 1.1083\!\times\!10^{-4} & 0
\end{bmatrix}$}.
\end{equation}
Its determinant satisfies
\begin{equation}
 \frac{17935}{10^{22}}<\det Y<\frac{17936}{10^{22}}.
\end{equation}
Define
\begin{equation}
 K(\boldsymbol{x}_0,\mathcal X):=\boldsymbol{x}_0-Y\boldsymbol F(\boldsymbol{x}_0)
 +\bigl(I_4-Y[D\boldsymbol F](\mathcal X)\bigr)(\mathcal X-\boldsymbol{x}_0).
\end{equation}
Let $\operatorname{diag}$ form a diagonal matrix, interpret multiplication of a diagonal matrix and a box componentwise, and let $\operatorname{int}\mathcal X$ denote the interior of $\mathcal X$. Set
\begin{equation}
 R_K:=\operatorname{diag}\!\left(\frac1{10},\frac1{20},\frac{41}{100},\frac{29}{50}\right).
\end{equation}
Exact rational interval evaluation gives
\begin{equation}
 K(\boldsymbol{x}_0,\mathcal X)
 \subset \boldsymbol{x}_0+R_K(\mathcal X-\boldsymbol{x}_0)
 \subset\operatorname{int}\mathcal X.
\end{equation}
The Krawczyk theorem gives existence and uniqueness of a zero in $\mathcal X$ \citep{Krawczyk1969,Moore1977}. Since $\boldsymbol F$ has rational polynomial entries and the zero is isolated, all four coordinates are real algebraic.
\end{proof}

\subsection{Global optimality of the algebraic candidate}

Tensor-product Bernstein evaluation on $I_{\tau}\times[-1/5,2]$ gives
\begin{equation}
\begin{aligned}
&M_1(\tau,\beta)>\tfrac14,\quad
 M_2(\tau,\beta)>\tfrac23,\quad
 M_3(\tau,\beta)>3,\\
&M_4(\tau,\beta)>25,\quad
 M_5(\tau,\beta)>26,\quad
 M_6(\tau,\beta)>300,\\
&M_7(\tau,\beta)>300,\quad
 M_8(\tau,\beta)>9000,
\end{aligned}
\label{eq:hurwitz_minor_bounds}
\end{equation}
for every $(\tau,\beta)\in I_\tau\times[-1/5,2]$.
Since the leading coefficient of $\delta_{\beta}$ is $\tau^2>0$, these inequalities place $I_{\tau}$ inside $\mathcal R_{\rm stab}$.

Let $\widehat{\mathbb K}:=\mathbb K(\widehat\eta_1,\widehat\eta_2,\widehat\tau,\widehat g)$. At $(\widehat\tau,\widehat g)$, exact division in $\widehat{\mathbb K}[\eta]$ gives
\begin{equation}
 H_-(\eta,\widehat\tau,\widehat g)
 =(\eta-\widehat\eta_1)^2(\eta-\widehat\eta_2)^2Q_8(\eta),
 \label{eq:Hminus_factor_compact}
\end{equation}
where $Q_8\in\widehat{\mathbb K}[\eta]$ has degree eight and the division has zero remainder. For general $\beta$, write
\begin{equation}
 H_{\beta}(\eta,\widehat\tau,\widehat g)
 =a_2(\eta)\beta^2+a_1(\eta)\beta+a_0(\eta)
 \label{eq:Hbeta_quadratic_compact}
\end{equation}
and define
\begin{equation}
\begin{aligned}
 q_L(\eta)&:=a_1(\eta)-\frac25a_2(\eta),\qquad
 q_U(\eta):=a_1(\eta)+4a_2(\eta),\\
 \Xi(\eta)&:=4a_2(\eta)a_0(\eta)-a_1(\eta)^2.
\end{aligned}
\end{equation}
Exact polynomial division gives
\begin{equation}
 \Xi(\eta)=\eta\,\widetilde\Xi(\eta),
 \qquad \widetilde\Xi\in\widehat{\mathbb K}[\eta].
\end{equation}
The sign calculations use rational interval arithmetic over
$\mathcal X$ and the Bernstein coefficients in
\eqref{eq:bernstein_coefficients}. For a polynomial $\Pi$ of degree
$n_{\Pi}$ on $[0,\infty)$, set
$\varrho_0:=\eta/(1+\eta)\in[0,1)$. After substituting
$\eta=\varrho_0/(1-\varrho_0)$ and multiplying by
$(1-\varrho_0)^{n_{\Pi}}$, the transformed polynomial extends
continuously to $\varrho_0=1$, so its Bernstein coefficients are
evaluated on $[0,1]$. Finite intervals use their affine maps to
$[0,1]$. For $[20,\infty)$, set
$\varrho_{\infty}:=1/\eta\in(0,1/20]$. After multiplication by the
positive power of $\varrho_{\infty}$ that clears the denominator,
the transformed polynomial extends continuously to
$\varrho_{\infty}=0$, so its sign is established on $[0,1/20]$
\citep{Garloff1986}.
\begin{table}[H]
\centering
\caption{Rational Bernstein sign bounds at the algebraic candidate}
\label{tab:case_sign_conditions}
\scriptsize
\begin{tabular}{@{}lcc@{}}
\toprule
Quantity and range & Transform/degree & Min. coefficient\\
\midrule
$Q_8(\eta)$, $[0,\infty)$ & compact/$200$ & $>9/10^6$\\
$H_+(\eta,\widehat\tau,\widehat g)$, $[0,\infty)$ & compact/$400$ & $>9/10^6$\\
$\widetilde\Xi(\eta)$, $(0,97/100]$ & affine/$200$ & $>7000$\\
$q_L(\eta)$, $[97/100,43/10]$ & affine/$20$ & $>1$\\
$\Xi(\eta)$, $[43/10,20]$ & affine/$200$ & $>3\times10^{11}$\\
$-q_U(\eta)$, $[20,\infty)$ & reciprocal/$40$ & $>3/2000$\\
\bottomrule
\end{tabular}
\end{table}
For every row of Table~\ref{tab:case_sign_conditions}, every interval Bernstein coefficient has a lower endpoint above the reported rational bound.

The exact stability domain also gives
\begin{equation}
 \tau>\frac{69}{2000}
 \qquad\forall\tau\in\mathcal R_{\rm stab},
 \label{eq:stable_lower_bound}
\end{equation}
and the isolating interval satisfies
\begin{equation}
 I_{\tau}\subset\left(\frac{69}{2000},\frac{37}{1000}\right).
 \label{eq:Itau_rational_bounds}
\end{equation}
Let $z>0$ and $w\in(0,1/400]$ denote rightward and leftward perturbations of $\widehat\tau$, respectively. At the two candidate stationary frequencies, quartic expansion in the time constant gives
\begin{align}
 H_-(\widehat\eta_1,\widehat\tau+z,\widehat g)
 &=z(\kappa_1+\kappa_2z+\kappa_3z^2+\kappa_4z^3),\label{eq:right_quartic}\\
 H_-(\widehat\eta_2,\widehat\tau-w,\widehat g)
 &=w(\ell_1+\ell_2w+\ell_3w^2+\ell_4w^3).\label{eq:left_quartic}
\end{align}
Exact coefficient enclosure over $\mathcal X$ gives
\begin{equation}
\begin{gathered}
-1527<\kappa_1<-1526,\quad -13916<\kappa_2<-13915,\\
-445<\kappa_3<-444,\quad -635<\kappa_4<-634,
\end{gathered}
\label{eq:kappa_signs}
\end{equation}
and
\begin{equation}
\begin{aligned}
-3252219&<\ell_1<-3252218,\\
1294048687&<\ell_2<1294048688,\\
200819160&<\ell_3<200819161,\\
229809521&<\ell_4<229809522.
\end{aligned}
\end{equation}
Thus \eqref{eq:right_quartic} is negative for every $z>0$. For $0<w\le1/400$, the bracket in \eqref{eq:left_quartic} is bounded above by
\begin{equation}
\begin{aligned}
&-3252218+\frac{1294048688}{400}
 +\frac{200819161}{400^2}\\
&\qquad +\frac{229809522}{400^3}
 =-\frac{506802223039}{32000000}<0.
\end{aligned}
\label{eq:left_exact_bound}
\end{equation}

\begin{theorem}[Unique global optimum]
\label{thm:global_optimum}
For the uncertain DOB in \eqref{eq:example_CQ}--\eqref{eq:example_P},
\begin{equation}
 \mathcal T_{\rm ex}^{\star}=\{\tau^{\star}\},
 \qquad \tau^{\star}=\widehat\tau\approx0.0364075,
\end{equation}
and
\begin{equation}
 \min_{\tau\in\mathcal R_{\rm stab}}
 \sup_{\beta\in[-1/5,2]}\|E_{\beta}(\cdot;\tau)\|_{\infty}
 =\gamma^{\star}:=\sqrt{\widehat g}\approx2.29975.
\end{equation}
The worst case is attained at $\beta=-1/5$ and at the two positive frequencies in \eqref{eq:candidate_frequency_values}.
\end{theorem}

\begin{proof}
The bounds \eqref{eq:hurwitz_minor_bounds} establish robust stability of the algebraic candidate. Equation~\eqref{eq:Hminus_factor_compact} and the first row of Table~\ref{tab:case_sign_conditions} give nonnegativity at $\beta=-1/5$, while the second row gives positivity at $\beta=2$. If $a_2(\eta)\le0$, the quadratic \eqref{eq:Hbeta_quadratic_compact} is minimized at an uncertainty endpoint. If $a_2(\eta)>0$, the signs of $q_L$ and $q_U$ identify an endpoint minimizer unless $q_L(\eta)<0<q_U(\eta)$; in that remaining case the interior minimum is $\Xi(\eta)/(4a_2(\eta))$. The last four rows of Table~\ref{tab:case_sign_conditions} cover every $\eta>0$. At $\eta=0$, exact DC recovery gives $N_{\beta}(0,\widehat\tau)=0$, while candidate stability gives $D_{\beta}(0,\widehat\tau)>0$. Consequently,
\begin{equation}
 H_{\beta}(\eta,\widehat\tau,\widehat g)\ge0
 \quad\forall\eta\ge0,\quad\forall\beta\in[-1/5,2].
\end{equation}
Equality holds at $(\beta,\eta)=(-1/5,\widehat\eta_1)$ and $(-1/5,\widehat\eta_2)$. Candidate stability gives
\[
 D_{-1/5}(\widehat\eta_i,\widehat\tau)>0,
 \qquad i=1,2,
\]
so these equalities are equivalent to
\[
 |E_{-1/5}(\mathrm j\sqrt{\widehat\eta_i};\widehat\tau)|^2=\widehat g,
 \qquad i=1,2.
\]
Consequently, $J_{\rm ex}(\widehat\tau)=\widehat g$.

For every robustly stable $\tau>\widehat\tau$, set $z=\tau-\widehat\tau>0$. Equations~\eqref{eq:right_quartic} and~\eqref{eq:kappa_signs} give $H_-(\widehat\eta_1,\tau,\widehat g)<0$. Since $D_{-1/5}(\widehat\eta_1,\tau)>0$ on $\mathcal R_{\rm stab}$, this is equivalent to
\[
 |E_{-1/5}(\mathrm j\sqrt{\widehat\eta_1};\tau)|^2>\widehat g.
\]
For every robustly stable $\tau<\widehat\tau$, \eqref{eq:stable_lower_bound} and~\eqref{eq:Itau_rational_bounds} give $0<\widehat\tau-\tau<1/400$. Equations~\eqref{eq:left_quartic}--\eqref{eq:left_exact_bound} then give $H_-(\widehat\eta_2,\tau,\widehat g)<0$. Since $D_{-1/5}(\widehat\eta_2,\tau)>0$, it follows that
\[
 |E_{-1/5}(\mathrm j\sqrt{\widehat\eta_2};\tau)|^2>\widehat g.
\]
Hence every other robustly stable time constant has a larger objective value.
\end{proof}

\begin{samepage}
\begin{corollary}[Lower-boundary separation]
\label{cor:interior_optimum}
For the DOB in \eqref{eq:example_CQ}--\eqref{eq:example_P},
$\tau_1^{\rm s}<\tau^{\star}<\tau_2^{\rm s}$, and
$J_{\rm ex}(\tau)>J_{\rm ex}(\tau^{\star})$ for every
$\tau\in(\tau_1^{\rm s},\tau^{\star})$.
\end{corollary}
\begin{proof}
Since $\tau^{\star}=\widehat\tau\in I_\tau$, the rational intervals in \eqref{eq:stability_endpoint_boxes} and \eqref{eq:Itau_exact} give
$\tau_1^{\rm s}<\tau^{\star}<\tau_2^{\rm s}$. The strict objective inequality follows from the $\tau<\tau^{\star}$ part of the proof of Theorem~\ref{thm:global_optimum}.
\end{proof}
\end{samepage}

\subsection{Design interpretation}

\begin{figure}[t]
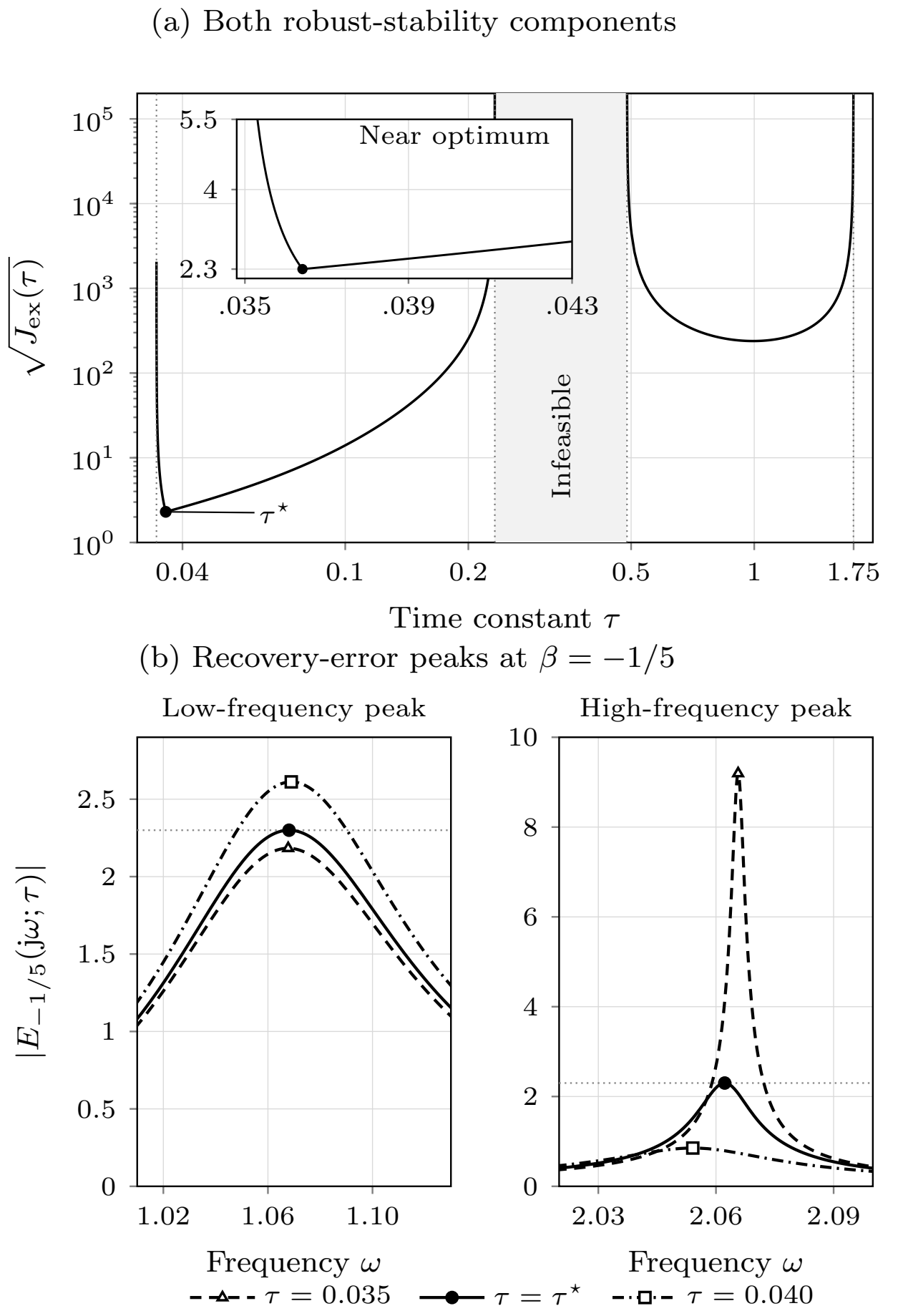

\centering

\caption{Time-constant selection: (a) worst-case recovery norm; (b) recovery-error frequency responses.}
\label{fig:case_design_meaning}
\end{figure}

Figure~\ref{fig:case_design_meaning} illustrates the design criterion in \eqref{eq:robust_recovery_value}--\eqref{eq:optimal_time_constants}. Panel~(a) plots the numerically evaluated norm with $W=1$ over both algebraically established robust-stability components; the inset resolves $\tau^{\star}$ in the first component. At $\tau=0.035$, which lies inside the first robust-stability component, the numerically evaluated worst-case norm is approximately $9.1945$. At $\tau=\tau^{\star}\approx0.0364075$, the two active peaks have the common value $2.29975$. Increasing the time constant to $0.04$ gives a worst-case norm of approximately $2.6118$. The uncertainty maximization used to draw panel (a) evaluates both endpoints and any interior stationary point of the squared frequency response in $\beta$, followed by numerical frequency maximization.

Panel~(b) magnifies the low- and high-frequency peaks in separate windows with different vertical ranges. The dotted horizontal lines mark the optimal norm, 2.29975. At the fixed uncertainty value $\beta=-1/5$, increasing $\tau$ from $0.035$ to $\tau^{\star}$ lowers the higher-frequency peak and raises the lower-frequency peak until their heights are equal. At $\tau=0.04$, the lower-frequency peak dominates. This comparison shows how the recovery criterion balances the two peaks within the prescribed filter family. Theorem~\ref{thm:global_optimum} establishes that the resulting time constant minimizes the objective over both robust-stability components.

\section{Discussion}

\subsection{What the procedure computes}

The semialgebraic solution yields the global performance level, the attainment status, and all minimizing time constants over the robust-stability components. In this case study, the optimizer is unique. For the present $Q$-filter, $|D_Q(\mathrm j\omega;\tau)-N_Q(\mathrm j\omega;\tau)|^2=\tau^2\omega^2(25+\tau^2\omega^2)$, which is strictly increasing in $\tau>0$ at every fixed nonzero frequency. Thus reducing $\tau$ improves this finite-frequency recovery factor. The global $H_{\infty}$ objective also reflects the uncertain closed-loop denominator $\delta$; Corollary~\ref{cor:interior_optimum} shows that every feasible $\tau$ between the lower boundary of the first robust-stability component and $\tau^{\star}$ has a larger worst-case norm. Hence the finite-frequency recovery intuition and the global robust-norm optimum are distinct.

\subsection{Additional performance channels}

A different scalar channel, such as a disturbance-to-output, noise-to-output, or control-effort transfer function, can be treated by the same machinery when its numerator and denominator depend affinely on the same uncertainty vector, its denominator is robustly Hurwitz with fixed degree and nonzero leading coefficient, and the channel is strictly proper. A finite collection of scalar channels can be enforced at prescribed levels by intersecting their feasible epigraph conditions. Their maximum can also be minimized by introducing a common epigraph variable.

\subsection{Computational scope}

The direct semialgebraic formulation contains $p+1$ quantified variables, whereas \eqref{eq:dimension_reduction} leaves only $\eta$ universally quantified. The number of edge conditions grows with the distinct affine corner images, and the polynomial degrees grow with plant, controller, weight, and $Q$-filter order. The method is particularly suited to low- and moderate-order SISO DOBs for which a global optimizer and a proof of global optimality are both valuable.

\section{Conclusion}

For a fixed-shape DOB $Q$-filter, a finite real-algebraic procedure computes the global nominal-response-recovery level, decides attainment, and, when attained, constructs all minimizing time constants over a possibly disconnected robust-stability domain. Recovery-error factorization, phase lifting, polynomial-edge reduction, and analytic edge-coordinate elimination yield a one-frequency semialgebraic predicate. The case study gives $\tau^{\star}\approx0.0364075$ and an interior robust-$H_{\infty}$ optimum separated from the lower boundary of the first robust-stability component.

\end{document}